\documentclass{appolb}

\newcommand{\be}{\begin{equation}}
\newcommand{\ee}{\end{equation}}
\newcommand{\ba}{\begin{eqnarray}}
\newcommand{\ea}{\end{eqnarray}}
\newcommand{\ban}{\begin{eqnarray*}}
\newcommand{\ean}{\end{eqnarray*}}
\newcommand \nn {\nonumber}

\usepackage{amssymb}
	
\begin{document}

\title{On the Dynamics of Unstable Quark-Gluon Plasma\thanks{Presented 
at the workshop of {\it ExtreMe Matter Institute} (EMMI) and XXVI 
Max-Born Symposium `Three Days of Strong Interactions', Wroc\l aw, 
Poland, July 9-11, 2009.}}

\author{Stanis\l aw Mr\' owczy\' nski
\address{Institute of Physics, Jan Kochanowski University, \\
ul.~\'Swi\c etokrzyska 15, 25-406 Kielce, Poland \\
and So\l tan Institute for Nuclear Studies, \\
ul.~Ho\.za 69, 00-681 Warsaw, Poland}}

\date{October 30, 209}

\maketitle

\begin{abstract}

Since the quark-gluon plasma, which is unstable due to 
anisotropic momentum distribution, evolves fast in time,
plasma's characteristics have to be studied as initial 
value problems. The chromodynamic fluctuations and the 
momentum broadening of a fast parton traversing the plasma 
are discussed here. The two quantities are shown to 
exponentially grow in time.

\end{abstract}


  
\section{Introduction}


The quark-gluon plasma (QGP), which is produced at the early stage 
of relativistic heavy-ion collisions, is most probably unstable 
due to anisotropic momentum distribution of quarks and gluons 
(partons), see the review \cite{Mrowczynski:2005ki}. The instability 
makes the system's state strongly time dependent, and consequently 
various plasma characteristics must be found as solutions of initial 
value problem. Two such characteristics are discussed in my lecture. 
I start with the fluctuations of chromodynamic fields and then, the 
correlation functions of the fields are used to compute the momentum 
broadening of a fast parton traversing the plasma. The two
characteristics exponentially grow in time due to unstable modes. 
Throughout my lecture, which is based on two recent publications
\cite{Mrowczynski:2008ae,Majumder:2009cf}, the plasma is assumed to
be weakly coupled.


\section{Chromodynamic fluctuations}


In the quark-gluon plasma (QGP), which is on average locally colorless,
chromodynamic fields, color charges and currents experience random 
fluctuations. In the equilibrium plasmas there are characteristic 
stationary spectra of fluctuations which can be found by means of 
the fluctuation-dissipation relations. Fluctuations in nonequilibrium 
systems evolve in time, and their characteristics usually depend on 
an initial state of the system. 

Fluctuations can be theoretically studied by means of several methods 
reviewed in the classical monographs \cite{Akh75,Sit82}. The method, 
which was chosen in \cite{Mrowczynski:2008ae}, is clearly exposed in 
the handbook \cite{LP81}.  It is applicable to both equilibrium and nonequilibrium plasmas but the initial plasma state is assumed to be 
on average charge neutral, stationary and homogeneous. It will be 
shown that when the plasma state is stable, the initial fluctuations 
exponentially decay and in the long time limit one finds a stationary 
spectrum of the fluctuations. When the initial state is unstable, the 
memory of initial fluctuations is not lost, as the unstable modes, 
which are usually present in the initial fluctuation spectrum, 
exponentially grow. 

The chromodynamic fluctuations are studied here using the transport 
theory of weakly coupled quark-gluon plasma which is formulated 
in terms of particles and classical fields. The particles - quarks, 
antiquarks and gluons - should be understood as sufficiently hard 
quasiparticle excitations of QCD quantum fields while the classical 
fields are highly populated soft gluonic modes. The transport equation 
of quarks reads 
\be
\label{transport-eq}
\big(D^0 + {\bf v} \cdot {\bf D} \big) Q(t,{\bf r},{\bf p})
- {g \over 2}
\{{\bf E}(t,{\bf r}) + {\bf v} \times {\bf B}(t,{\bf r}), 
\nabla_p Q(t,{\bf r},{\bf p}) \}
= 0 \;, 
\ee
where $Q(t,{\bf r},{\bf p})$ is the on-mass-shell quark distribution 
function which is $N_c\times N_c$ hermitean matrices belonging to 
the fundamental representation of the SU($N_c$) group; the covariant 
derivative in the four-vector notation reads 
$D^\mu \equiv \partial^\mu - ig [A^\mu(x),\cdots \;]$ and 
${\bf E}(t,{\bf r})$ and ${\bf B}(t,{\bf r})$ are the chromoelectric 
and chromomagnetic fields. The symbol $\{\dots , \dots \}$ denotes 
the anticommutator. Since the fluctuations of interest are assumed 
to be of the time scale, which is much shorter than that of inter-parton 
collisions, the collision terms are absent in Eq.~(\ref{transport-eq}). 
There are analogous transport equations for antiquark 
($\bar Q(t,{\bf r},{\bf p})$) and gluon ($G(t,{\bf r},{\bf p})$) 
distribution functions.

The transport equations are supplemented by the Yang-Mills equations 
describing a self-consistent generation of the chromoelectric and 
chromomagnetic fields by the color four-current $j^\mu =(\rho , {\bf j})$ 
$$
j^\mu_a (t,{\bf r}) = - g \int {d^3 p \over (2\pi)^3} \,
\frac{p^\mu}{E_{\bf p}}
{\rm Tr}\Big[\tau^a \big(Q (t,{\bf r},{\bf p})
- \bar Q(t,{\bf r},{\bf p}) \big)
+ T^a G(t,{\bf r},{\bf p}) \Big] \;,
$$
where $\tau^a$, $T^a$ with $a = 1, ... \, ,N_c^2-1$ are the SU($N_c$)
group generators in the fundamental and adjoint representations. 

I consider small deviations from a stationary homogeneous 
state which is globally and locally colorless; there are no currents 
as well. The quark distribution function of this state is 
$Q^0_{nm}({\bf p}) = n ({\bf p}) \: \delta^{nm}$. Due to the 
absence of color charges and currents in the stationary and 
homogeneous state, the chromoelectric ${\bf E}(t,{\bf r})$ and 
chromomagnetic ${\bf B}(t,{\bf r})$ fields are expected to vanish
while the potentials $A^0(t,{\bf r}), {\bf A}(t,{\bf r})$ are of 
pure gauge only. Since the plasma under considerations is assumed
to be weakly coupled with the perturbative vacuum state, the 
potentials can be gauged away to vanish. 

The quark distribution function is written down as
$Q(t,{\bf r},{\bf p}) =  Q^0({\bf p}) + 
\delta Q(t,{\bf r},{\bf p})$, and we assume that
$|Q^0| \gg |\delta Q| $ and $|\nabla_p Q^0| \gg | \nabla_p \delta Q|$
with the analogous formulas for antiquarks and gluons.
The transport (\ref{transport-eq}) and Yang-Mills equations 
are linearized in deviations from the stationary homogeneous 
state. We assume that $\delta Q$, ${\bf E}$, ${\bf B}$, $A^0$ 
and ${\bf A}$ are all of the same order. The linearized transport 
equation is
$$
\Big(\frac{\partial}{\partial t} + {\bf v} \cdot \nabla \Big) 
\delta Q(t,{\bf r},{\bf p})
- g \big({\bf E}(t,{\bf r}) + {\bf v} \times {\bf B}(t,{\bf r})\big) 
\nabla_p n({\bf p})
= 0 \;. 
$$
After the linearization the Yang-Mills equations get the familiar 
form of Maxwell equations of multi-component electrodynamics. 

The linearized transport and Maxwell equations are solved
with the initial conditions 
$\delta Q(t\!=\!0,{\bf r},{\bf p}) = \delta Q_0({\bf r},{\bf p})$,
${\bf E}(t\!=\!0,{\bf r}) = {\bf E}_0({\bf r})$ and \break
${\bf B}(t\!=\!0,{\bf r})= {\bf B}_0({\bf r})$,
by means of the one-sided Fourier transformation defined as
$$
f(\omega,{\bf k}) = \int_0^\infty dt \int d^3r 
e^{i(\omega t - {\bf k}\cdot {\bf r})}
f(t,{\bf r}) \;.
$$
The chromoelectric field, which solves the equations, is found as
\ba
\label{E-field2} 
\big[ - {\bf k}^2 \delta^{ij} &+& k^ik^j 
+ \omega^2 \varepsilon^{ij}(\omega,{\bf k}) \big] E^j_a(\omega,{\bf k})
= 
\\[2mm] \nn
&-& i\frac{g^2}{2} \int {d^3p \over (2\pi)^3} \,
\frac{v^i \big({\bf v}\times {\bf B}_{a0}({\bf k})\big)^j
\nabla_p^j f({\bf p})}{\omega - {\bf v}\cdot {\bf k}}
\\[2mm] \nn
&-& g \omega \int {d^3p \over (2\pi)^3} \,
\frac{v^i} {\omega - {\bf k}\cdot {\bf v}}\, 
\delta N^a_0({\bf k},{\bf p})
+ i \omega E_{a0}^i({\bf k})
-i \big({\bf k} \times {\bf B}_{a0}({\bf k})\big)^i \;,
\ea
where $f({\bf p}) \equiv n({\bf p}) + \bar n ({\bf p})
+ 2N_c n_g({\bf p})$ and $\delta N^a_0({\bf r},{\bf p}) \equiv
{\rm Tr}\big[\tau^a \big(\delta Q_0 ({\bf r},{\bf p})
- \delta \bar Q_0({\bf r},{\bf p}) \big)
+ T^a \delta G_0({\bf r},{\bf p}) \big]$;
$\varepsilon^{ij}(\omega,{\bf k})$ is the chromodielectric 
tensor of, in general, anisotropic plasma in the collisionless limit;
$\varepsilon^{ij}(\omega,{\bf k})$ does not carry any color indices,
as it corresponds to a colorless state of the plasma.

When the plasma stationary state is isotropic, the dielectric tensor 
can be expressed through its longitudinal ($\varepsilon_L(\omega,{\bf k})$)
and transverse ($\varepsilon_T(\omega,{\bf k})$) components and the matrix $\Sigma^{ij}(\omega,{\bf k}) 
\equiv - {\bf k}^2 \delta^{ij} + k^ik^j
+ \omega^2 \varepsilon^{ij}(\omega,{\bf k})$ 
from the left-hand-side of Eq.~(\ref{E-field2}) 
can be inverted as 
$$
 (\Sigma^{-1})^{ij}(\omega,{\bf k}) = 
\frac{1}{\omega^2 \varepsilon_L(\omega,{\bf k})}
\frac{k^ik^j}{{\bf k}^2}
+ \frac{1}{\omega^2 \varepsilon_T(\omega,{\bf k})-{\bf k}^2}
\Big(\delta^{ij} - \frac{k^ik^j}{{\bf k}^2}\Big) \;.
$$
Then, Eq.~(\ref{E-field2}) provides an explicit expression of
the chromoelectric field. Using the Maxwell equations, the chromomagnetic 
field, color current and color density can be all expressed through 
the chromoelectric field. 

The correlation functions 
$\langle E^i_a(t_1,{\bf r}_1) E^j_b(t_2,{\bf r}_2) \rangle$,
$\langle B^i_a(t_1,{\bf r}_1) B^j_b(t_2,{\bf r}_2) \rangle$,
where $\langle \cdots \rangle$ denotes averaging over statistical
ensemble, are determined by the initial correlations such as
$\langle \delta N_0^a({\bf r}_1,{\bf p}_1) 
\delta N_0^b({\bf r}_2,{\bf p}_2)\rangle$, 
$\langle E_{a0}^i({\bf r}_1) E_{b0}^j({\bf r}_2) \rangle$,
$\langle \delta N_0^a({\bf r}_1,{\bf p}_1) 
E_{b0}^j({\bf r}_2) \rangle$ which can be all expressed, 
using the Maxwell equations, through the correlation function 
of the distribution functions. The latter one is identified with
the respective correlation function of the classical system of 
free quarks, antiquarks and gluons which on average is stationary 
and homogeneous. For quarks the free correlation function is
\ban
\langle \delta Q^{mn}(t_1,{\bf r}_1,{\bf p}_1) 
\delta Q^{pr}(t_2,{\bf r}_2,{\bf p}_2)\rangle_{\rm free}
&=& \delta^{mr} \delta^{np} 
(2\pi )^3 \delta^{(3)}({\bf p}_1 - {\bf p}_2) \,
\\[2mm] \nonumber
&\times&
\delta^{(3)}\big({\bf r}_2 - {\bf r}_1 
- {\bf v}_1(t_2 - t_1)\big) \: n({\bf p}_1) \;.
\ean
 
In the case of equilibrium plasma, where all collective modes
are damped, I consider the times which are much longer than 
the decay time of collective excitations. Then, the correlation 
function of the chromoelectric fields equals
$$
\langle E_a^i(t_1,{\bf r}_1) E_b^j(t_2,{\bf r}_2) \rangle_\infty 
= 
\int {d\omega \over 2\pi} {d^3k \over (2\pi)^3}
e^{-i \big(\omega (t_1 - t_2)
 - {\bf k}\cdot ({\bf r}_1 - {\bf r}_2)\big)} 
\langle E_a^i E_b^j\rangle_k \;,
$$ 
where the fluctuation spectrum is
$$
\langle E_a^i E_b^j\rangle_k
= 
2 \delta^{ab} T \omega^3 \bigg[
\frac{k^ik^j}{{\bf k}^2}
\frac{\Im \varepsilon_L(\omega,{\bf k})}
{|\omega^2 \varepsilon_L(\omega,{\bf k})|^2}
+
\Big(\delta^{ij} - \frac{k^ik^j}{{\bf k}^2}\Big)
\frac{\Im \varepsilon_T(\omega,{\bf k})}
{|\omega^2 \varepsilon_T(\omega,{\bf k})-{\bf k}^2|^2}
\bigg] \;.
$$
As seen, the fluctuation spectrum has strong peaks corresponding
to the collective modes determined by the equations
$\varepsilon_L(\omega,{\bf k}) = 0$
and $\omega^2 \varepsilon_T(\omega,{\bf k}) - {\bf k}^2 = 0$.

As an example of a nonequilibrium situation, I discuss 
fluctuations of longitudinal chromoelectric fields in the 
two-stream  system which is unstable with respect to  
longitudinal modes. Nonequlibrium calculations are much more 
difficult than the equilibrium ones. The first problem is to 
invert the matrix $\Sigma^{ij}(\omega,{\bf k})$. In the case 
of longitudinal electric field, which is discussed here, the 
matrix is replaced by the scalar function. 

The distribution function of the two-stream system is chosen to be 
$$
f({\bf p}) = (2\pi )^3 n 
\Big[\delta^{(3)}({\bf p} - {\bf q}) + \delta^{(3)}({\bf p} 
+ {\bf q}) \Big] \;,
$$
where $n$ is the effective parton density in a single stream. 
There are four roots $\pm \omega_{\pm}({\bf k})$ of the dispersion 
equation $\varepsilon_L(\omega,{\bf k}) = 0$. The solution 
$\omega_+({\bf k})$ represents the stable modes and $\omega_-({\bf k})$
corresponds to the well-known two-stream electrostatic instability
for ${\bf k} \cdot {\bf u} \not= 0$ and 
${\bf k}^2 ({\bf k} \cdot {\bf u})^2 < 2 \mu^2 
\big({\bf k}^2 - ({\bf k} \cdot {\bf u})^2\big)$
where ${\bf u} \equiv {\bf q}/|{\bf q}|$ is the stream velocity
and $\mu^2 \equiv g^2n/2|{\bf q}|$. Then, 
$\omega_-({\bf k})=i\gamma_{\bf k}$ with $0 \le \gamma_{\bf k} \in R$.

The correlation function of longitudinal chromoelectric
fields generated by the unstable modes is found as 
\ba
\label{E^iE^j-2-stream}
&& \langle E_a^i(t_1,{\bf r}_1) E_b^j(t_2,{\bf r}_2) 
\rangle_{\rm unstable}
=
\\[2mm] \nn
&& \frac{g^2}{2}\,\delta^{ab} n 
\int {d^3k \over (2\pi)^3} 
\frac{e^{i {\bf k}({\bf r}_1 - {\bf r}_2)}}{{\bf k}^4}
\frac{k^i k^j\big(\gamma_{\bf k}^2 + ({\bf k} \cdot {\bf u})^2\big)^2}
{(\omega_+^2 - \omega_-^2)^2\gamma_{\bf k}^2}
\\[2mm] \nn 
&&\times
\Big[
\big(\gamma_{\bf k}^2 + ({\bf k} \cdot {\bf u})^2\big)
\cosh \big(\gamma_{\bf k} (t_1 + t_2)\big)
+
\big(\gamma_{\bf k}^2 - ({\bf k} \cdot {\bf u})^2\big)
\cosh \big(\gamma_{\bf k} (t_1 - t_2)\big) \Big] \,.
\ea
As seen, the correlation function of the unstable system is
invariant with respect to space translations -- it depends on the 
difference $({\bf r}_1 - {\bf r}_2)$ only. The plasma state, which 
is initially on average homogeneous, remains like this in course of
the system's temporal evolution. The time dependence of the correlation 
function is very different from the space dependence. The electric 
fields exponentially grow and so does the correlation function both 
in $(t_1 + t_2)$ and $(t_1 - t_2)$. The fluctuation spectrum also 
evolves in time, as the growth rate of unstable modes is wave-vector 
dependent and after a sufficiently long time the fluctuation spectrum 
is dominated by the fastest growing modes.


\section{Momentum broadening of a fast parton}


When a highly energetic parton travels through dense QCD matter, 
it receives random kicks from elastic interactions with constituents 
of the plasma. The average transverse momentum transfer per unit 
path length is related to the radiative energy loss of the parton 
\cite{Baier:1996sk}. The parameter describing the average amount 
of transverse momentum broadening per unit length is called 
$\hat{q}$ and is defined as
\be
\hat{q} \equiv d\langle\Delta {\bf p}_T^2\rangle / dz ,
\label{qhat}
\ee
when the fast parton flies along the direction $z$. The values 
of $\hat{q}$ extracted from experimental data on relativistic 
heavy-ion collisions vary in a rather broad range 
$0.5-15\;{\rm GeV^2/fm}$ depending on the model of hard particle 
propagation in strongly interacting matter produced in nuclear 
collisions \cite{Eskola:2004cr,Baier:2006fr}. 

The calculations of $\hat{q}$ for the case of perturbative 
quark-gluon plasma in equilibrium are well understood (see 
\cite{Arnold:2008vd,Peigne:2008wu} for recent work). For such 
a plasma the value of $\hat{q}$ is predicted to lie at the 
lower end of the range of values deduced from experiments
\cite{Baier:2006fr}. However, the plasma momentum distribution 
is initially anisotropic and recently $\hat{q}$ has been 
computed \cite{Romatschke:2006bb,Baier:2008js} for such a plasma.
However, the fact that the anisotropic plasma as an unstable system 
evolves fast in time has not been taken into account. It seems 
rather unjustified, as the numerical simulations 
\cite{Dumitru:2007rp,Schenke:2008gg} clearly indicate that 
$\hat{q}$ receives a sizable contribution from the unstable 
growing modes. 

An analytic approach to compute $\hat{q}$ in unstable plasma
has been developed in \cite{Majumder:2009cf}. It formulates the 
transverse momentum fluctuations in terms of classical Langevin 
problem. $\hat{q}$ is computed by treating the parton as an energetic 
classical particle with SU(3) color charge moving in the presence 
of the fluctuating color fields. Then, $\hat{q}$ is expressed 
through the correlation function of chromodynamic fields computed 
in \cite{Mrowczynski:2008ae}. For the equilibrium plasma the 
Langevin approach recovers the known result obtained within the 
standard thermal field theory \cite{lebellac}.

Let me consider a classical parton which moves across a quark-gluon 
plasma.  Its motion is described by the Wong equations 
\cite{Wong:1970fu}
\ba
\label{EOM-1a}
\frac{d x^\mu(\tau)}{d \tau} &=& u^\mu(\tau ) ,
\\ [2mm]
\label{EOM-1b}
\frac{d p^\mu(\tau)}{d \tau} &=&
g Q^a(\tau ) \, F_a^{\mu \nu}\big(x(\tau )\big) 
\, u_\nu(\tau ) ,
\\ [2mm]
\label{EOM-1c}
\frac{d Q_a(\tau)}{d \tau} &=& 
- g f^{abc} p_\mu (\tau ) \, 
A^\mu _b \big(x(\tau )\big) \, 
Q_c(\tau) ,
\ea
where $\tau$, $x^\mu(\tau )$, $u^\mu(\tau)$ and  $p^\mu(\tau)$
are, respectively, the parton's  proper time, its trajectory, 
four-velocity and  four-momentum; $F_a^{\mu \nu}$ and  
$A_a^\mu$ denote the chromodynamic field strength tensor and
four-potential, respectively, and $Q^a$ is the classical 
color charge of the parton.

We look for a solution of the Wong equations in a specific
gauge assuming that the potential vanishes along the parton's 
trajectory {\it i.e.} our gauge condition is
$$
p_\mu (\tau ) \, A^\mu _a \big(x(\tau )\big) = 0 .
$$
Then, Eq.~(\ref{EOM-1c}) simply tells that $Q_a$ is constant
as a function of $\tau$. 

One solves Eqs.~(\ref{EOM-1a}, \ref{EOM-1b}) assuming that the 
parton's momentum ${\bf p}$ is so high that its changes 
$\Delta {\bf p}$ caused by the interactions with the medium 
are small compared with ${\bf p}$. The changes of the velocity 
vector ${\bf v}$ are then negligible, and we can consider the 
parton to move along a straight-line path with constant velocity. 

Assuming that the parton moves with the speed of light in 
the {\em positive} $z$-direction that is $x^\mu(t) = (t,0,0,t)$, 
one finds
\be
\label{EOM-4}
p^\mu(t) = p^\mu(0) + g Q_a \int_0^t dt' 
\left[F_a^{\mu 0}(t') - F_a^{\mu 3}(t) \right] ,
\ee
where $F_a^{\mu \nu}(t)$ should be understood as a short hand
notation of $F_a^{\mu \nu}\big(x(t)\big)$.

In the spirit of Langevin approach, we consider the ensemble 
average $\langle p^\mu(t) p^\nu(t) \rangle$ indicated by
the angular brackets. The ensemble average involves averaging
over color charges which is performed by means of the relation
$$
\int dQ \,Q_a Q_b = C_2 \delta^{ab},
$$
where $C_2 = 1/2$ for particles (quarks) in fundamental 
representation of the ${\rm SU}(N_c)$ group and $C_2 = N_c$ 
for particles (gluons) in adjoint representation. Then, we find
\ba
\label{p^mu-p^nu-1}
&&\langle p^\mu(t) p^\nu(t) \rangle = 
\langle p^\mu(0) p^\nu(0) \rangle
\\ [2mm] \nn 
&&+ g^2 \frac{C_R}{N_c^2 -1} 
\int_0^t dt_1 \int_0^t dt_2
\langle \big(F_a^{\mu 0}( t_1) - F_a^{\mu 3}(t_1) \big)
\big(F_a^{\nu 0}(t_2) - F_a^{\mu 3}(t_2) \big)
\rangle \;,
\ea
where $C_R$ ($R=F,A$) is the eigenvalue of the quadratic 
Casimir operator, $C_F = (N_c^2-1)/2N_c$ and
$C_A = N_c$.

Introducing $\Delta{\bf p}_T(t) \equiv 
{\bf p}_T(t) - {\bf p}_T(0)$
and $\langle \Delta {\bf p}_T^2(t)\rangle \equiv
\langle \Delta {\bf p}_T(t) \cdot 
\Delta {\bf p}_T(t) \rangle$, we have
\ba
\nn
\langle \Delta {\bf p}_T^2(t)\rangle 
&=& g^2 \frac{C_R}{N_c^2 -1}
 \int_0^t dt_1  \int_0^t dt_2 
\bigg[
    \langle E^x_a(t_1) E^x_a(t_2)\rangle
 +  \langle E^y_a(t_1) E^y_a(t_2)\rangle
\\ \label{dp_T-dp_T-1}
&-&  \langle E^x_a(t_1) B^y_a(t_2)\rangle
 +   \langle E^y_a(t_1) B^x_a(t_2)\rangle
 -  \langle B^y_a(t_1) E^x_a(t_2)\rangle
\\ \nn
&+&  \langle B^x_a(t_1) E^y_a(t_2)\rangle
 +   \langle B^x_a(t_1) B^x_a(t_2)\rangle
 +   \langle B^y_a(t_1) B^y_a(t_2)\rangle 
\bigg]\,,
\ea
where, say, $E^x_a(t)$ should be understood as
$E^x_a\big(t,{\bf r}(t)\big)$ with 
${\bf r}(t) \equiv (0,0,t)$ that is only the fields at 
the parton's trajectory enter Eq.~(\ref{dp_T-dp_T-1}). 

Eq.~(\ref{dp_T-dp_T-1}) has been derived in the specific gauge 
and the right-hand side of the equation is, in general, 
gauge dependent. However, the field correlation functions 
derived in \cite{Mrowczynski:2008ae} are gauge independent 
within the Hard-Loop Approximation, as discussed in detail in 
Sec.~VIIIA of \cite{Mrowczynski:2008ae}. Therefore, 
Eq.~(\ref{dp_T-dp_T-1}) is gauge independent within the used 
approximations. 

Let us now assume that the quark-gluon plasma is translationally 
invariant in space and time and, hence, the field correlators 
depend only on the difference of the field's arguments. This 
assumption, which is relevant for equilibrium plasmas, will be 
not adopted for the two-stream system as the growth of unstable 
modes break the translational invariance in time. Making use of the translational invariance, one introduces the fluctuation spectrum
\be
\langle E_a^i E_a^j \rangle_k
\equiv \int dt \int d^3r e^{i(\omega t -{\bf k}{\bf r})} 
\langle E_a^i(t,{\bf r}) E_a^j(0,{\bf 0}) \rangle \,,
\ee
which allows us to write down Eq.~(\ref{dp_T-dp_T-1}) in the form
\ba
\label{dp_T^2-1}
\langle \Delta {\bf p}_T^2(t)\rangle
&=& g^2 
\frac{C_R}{N_c^2 -1}
\int_0^t dt_1 \int_0^t dt_2
\int \frac{d^4k}{(2\pi)^4} \; 
e^{i(\omega - k_z)(t_1 - t_2)}
\\ \nn
&\times&
\bigg[
\langle E^x_a E^x_a \rangle_k 
 +  \langle E^y_a E^y_a \rangle_k 
 -  \langle E^x_a B^y_a \rangle_k
 +  \langle E^y_a B^x_a \rangle_k 
\\ \nn
&&- \langle B^y_a E^x_a \rangle_k
 +  \langle B^x_a E^y_a \rangle_k
 +  \langle B^x_a B^x_a \rangle_k
 +  \langle B^y_a B^y_a \rangle_k 
\bigg]
\ea
Since the double integral over $t_1$ and $t_2$ tends 
to a delta function of $(\omega - k_z)/2$ in the long-time 
limit
$$
\int_0^t dt_1 \int_0^t dt_2
e^{i(\omega - k_z)(t_1 - t_2)}
= \frac{4 \sin\Big(\frac{(\omega - k_z) t}{2} \Big)}{(\omega - k_z)^2} 
\buildrel{t \rightarrow \infty}\over{\longrightarrow} 
\pi t \delta \Big(\frac{\omega - k_z}{2}\Big) ,
$$
we find the transport coefficient 
$\hat q = d\langle \Delta {\bf p}_T^2(t)\rangle /dt$ as 
\ba
\label{qhat-formula}
\hat q &=& 
2\pi g^2 \frac{C_R}{N_c^2 -1}
 \int \frac{d^4k}{(2\pi)^4} \; 
\delta(\omega - k_z) 
\bigg[\langle E^x_a E^x_a \rangle_k 
 +  \langle E^y_a E^y_a \rangle_k
\\ \nn 
&-&  \langle E^x_a B^y_a \rangle_k
 + \langle E^y_a B^x_a \rangle_k
 -  \langle B^y_a E^x_a \rangle_k
 +  \langle B^x_a E^y_a \rangle_k
 + \langle B^x_a B^x_a \rangle_k
 +  \langle B^y_a B^y_a \rangle_k 
\bigg] \,.
\ea
Using the equilibrium field correlators derived 
in \cite{Mrowczynski:2008ae}, Eq.~(\ref{qhat-formula}) gives
\ba
\label{qhat-class}
\hat q =
2 g^2 C_R T \int \frac{d^3k}{(2\pi)^3} \: 
\frac{k_T^2}{k_z {\bf k}^2} 
\left[ \frac{\Im \varepsilon_L(k_z,{\bf k})}
{|\varepsilon_L(k_z,{\bf k})|^2} 
+ \frac{ k_z^2 k_T^2 \: \Im \varepsilon_T(k_z,{\bf k})}
{| k_z^2 \varepsilon_T(k_z,{\bf k}) - {\bf k}^2|^2} \right] .
\ea
The classical formula (\ref{qhat-class}) holds for the inverse 
wave vectors of the fields which are much longer than the de 
Broglie wavelength of plasma particles. It requires 
$|{\bf k}| \ll T$ where $T$ is the plasma temperature. 
For larger wave vectors a quantum approach discussed 
in \cite{Majumder:2009cf} is needed. 

Let us now consider the momentum broadening of a fast parton
in unstable anisotropic plasmas. For the sake of analytical 
tractability, the two-stream plasma is discussed and only 
longitudinal electric fields are taken into account.
Substituting the correlation function (\ref{E^iE^j-2-stream}) 
into Eq.~(\ref{dp_T-dp_T-1}), one finds 
\ba
\nn
\langle \Delta {\bf p}_T^2(t)\rangle 
&=&
g^2 \frac{C_R}{N_c^2 -1}
\int_0^t dt_1 \int_0^t dt_2
\Big[
\langle E^x_a(t_1)\: E^x_a(t_2)\rangle
+ \langle E^y_a(t_1)\: E^y_a(t_2)\rangle \Big]
\\[2mm] \label{p2-2-stream-1}
&=&
\frac{g^4}{4} \, C_R \, n 
\int \frac{d^3k}{(2\pi)^3} 
\frac{k_T^2}{{\bf k}^4 (\omega_+^2 - \omega_-^2)^2}
\frac{\big(\gamma_{\bf k}^2 + ({\bf k} \cdot {\bf u})^2\big)^2} 
{\gamma_{\bf k}^2(k_z^2 + \gamma_{\bf k}^2)}
\\[2mm] \nn 
&\times& 
\bigg[
\big(\gamma_{\bf k}^2 + ({\bf k} \cdot {\bf u})^2\big)
\big(|e^{(i k_z + \gamma_{\bf k})t}-1|^2
   + |e^{(i k_z - \gamma_{\bf k})t}-1|^2 \big)
\\ [2mm] \nn 
&& + 4 \big(\gamma_{\bf k}^2 - ({\bf k} \cdot {\bf u})^2\big)
\frac{k_z^2 - \gamma_{\bf k}^2}{k_z^2 + \gamma_{\bf k}^2}
\bigg] .
\ea

When only the fastest growing mode is included, 
Eq.~(\ref{p2-2-stream-1}) changes to
\be
\label{qhat-unstable}
\hat{q} = \frac{d\langle {\bf p}_T^2(t)\rangle}{dt} 
 \approx 
\frac{g^4}{2} \, C_R \, n  
\int \frac{d^3k}{(2\pi)^3} \: e^{2\gamma_{\bf k}t} \:
\frac{k_T^2 \big(\gamma_{\bf k}^2 + ({\bf k} \cdot {\bf u})^2\big)^3}
{{\bf k}^4 (\omega_+^2 - \omega_-^2)^2
\gamma_{\bf k} (k_z^2 + \gamma_{\bf k}^2)} .
\ee
It should be noted that Eq.~(\ref{qhat-unstable}), in contrast 
to Eq.~(\ref{p2-2-stream-1}), suffers from divergence when 
$\gamma_{\bf k} \to 0$. The momentum broadening (\ref{qhat-unstable}) 
grows exponentially in time, as the exponentially growing fields 
exert an exponentially growing influence on the propagating 
parton. This effect is missing in the previous results 
\cite{Romatschke:2006bb,Baier:2008js} obtained for the anisotropic 
plasma treated as a stationary medium.


\section{Discussion and Outlook}


If the momentum distribution of partons is anisotropic, the 
quark-gluon is unstable with respect the chromomagnetic modes.
The instability growth rate $\gamma$ is of order $gT$ 
\cite{Mrowczynski:2005ki} where $T$ is here the characteristic 
parton momentum corresponding to temperature of equilibrium plasma. 
The time scale of unstable mode growth $\gamma^{-1}$ is the
shortest dynamical time scale in weakly coupled plasmas. 
Therefore, the unstable plasma cannot be treated as a static
medium whenever a characteristics, which involves color degrees 
of freedom, is studied. 

In this lecture I discussed the fluctuations of chromodynamic
fields in the plasma, and then, the field correlation functions
were used to compute the momentum broadening of a fast parton
flying across the plasma. Both quantities were found as solutions
of initial value problem. In the case of unstable plasma, the 
field fluctuations as well as the momentum broadening were
explicitly shown to exponentially grow in time.

An analytic treatment of unstable plasma is rather difficult. 
Therefore, instead of the plasma with momentum distribution 
relevant for relativistic heavy-ion collisions, a toy model 
representing the two stream system was discussed. The model 
grasps some important features of unstable systems, but it 
cannot be used for any quantitative predictions on the 
quark-gluon plasma produced in high-energy nucleus-nucleus 
collisions. An analysis the chromodynamic fluctuations and 
momentum broadening in such plasma is in progress but many 
other plasma characteristics, in particular transport 
coefficients, need to be studied to understand QGP from 
early stage of relativistic heavy-ion collisions. 


\end{document}